\documentclass[12pt]{article}
\usepackage{graphicx}
\usepackage{amsmath}
\usepackage{subfigure}
\usepackage{graphicx}
\usepackage{amsmath}
\usepackage{cite}

\begin{document}
\begin{center}
{\Large\bf Thermodynamics of the Variable Modified Chaplygin gas }\\[8 mm]
D. Panigrahi\footnote{ Sree Chaitanya College, Habra 743268, India
\emph{and also} Relativity and Cosmology Research Centre, Jadavpur
University, Kolkata - 700032, India , e-mail:
dibyendupanigrahi@yahoo.co.in }
  and S. Chatterjee\footnote{Relativity and Cosmology Research Centre,
Jadavpur University, Kolkata - 700032, India, e-mail : chat\_sujit1@yahoo.com} \\[10mm]

\end{center}
\begin{abstract}

A cosmological model with a new variant of Chaplygin gas obeying
an equation of state(EoS), $P = A\rho - \frac{B}{\rho^{\alpha}}$
where $B= B_{0}a^{n}$ is investigated in the context of its
thermodynamical behaviour. Here $B_{0}$ and
 $n$ are constants and $a$ is the scale factor. We show that the
equation of state of this `Variable Modified Chaplygin gas' (VMCG)
can describe the current accelerated expansion of the universe.
Following standard thermodynamical criteria we mainly discuss the
classical thermodynamical stability of the model and find that the
new parameter, $n$  introduced in VMCG plays a crucial role in
determining the stability considerations and should always be
\emph{negative.} We further observe that although the earlier
model of  Lu explains many of the current observational findings
of different probes it fails the desirable tests of
thermodynamical stability. We also note that for $n < 0$ our model
points to a phantom type of expansion which, however,   is found to be
compatible with current SNe Ia observations and CMB anisotropy
measurements. Further the third law of thermodynamics is obeyed in
our case. Our model is very general in the sense that many of
earlier works in this field may be obtained as a special case of
our solution. An interesting point to note is that the  model also
apparently suggests a smooth transition from the big bang to the
big rip in its whole evaluation process.
\end{abstract}

   ~~~~KEYWORDS : cosmology;Chaplygin gas;thermodynamics\\
   \vspace{-0.5cm }

~~\hspace{0.2 cm} PACS :   98.80.-k,98.80.Es,95.30.Tg,05.70.Ce

\section*{1. Introduction}

Following the high redshift supernovae data in the last decade
~\cite{rei, aman} we know that when interpreted within the
framework of the standard FRW type of universe (homogeneous and
isotropic) we are left with the only alternative that the universe
is now going through an accelerated expansion with baryonic matter
contributing only $5\%$ of the total budget. Later data from CMBR
studies ~\cite{aman} further corroborate this conclusion which has
led a vast chunk of cosmology community (~\cite{ben} and
references therein) to embark on a quest to explain the cause of
the acceleration. In fact the studies on accelerated expansion and
its possible interpretations from different angles have been
reigning the research paradigm for the last few decades. The
teething problem now confronting researchers in this field is the
identification of the mechanism that triggered the late inflation.
But as they are already discussed extensively in the literature
(we are sparing the readers here to repeat once again those
arguments) all
the alternatives face serious theoretical problems in some form or other.

In the literature  a good number of dark energy models are
proposed but little is precisely known about its origin. Nowadays, the
dark energy problem remains one of the major unsolved problems of
theoretical physics~\cite{wen}. On the  way of searching for
possible solutions of this problem various models are explored
during last few decades referring to e.g. new exotic forms of
matter (\emph{e.g.} quintessence) ~\cite{quint}, phantom
~\cite{phan}, holographic models ~\cite{hol}, string theory
landscape~\cite{string}, Born-Infeld quantum condensate
~\cite{quantum}, modified gravity approaches~\cite{star},
inhomogeneous spacetime~\cite{krasin}, higher dimensional space time~\cite{dp1} etc.

While the above mentioned alternatives to address the observed
acceleration of the current phase have both positive and negative
aspects a number of papers have come up taking into account the
Chaplygin gas~\cite{kam, dp,sc1} as a new form of matter field to
simulate unified dark energy and dark matter model. It is presumed
that for gravitational attraction, the dark matter component is
responsible for galaxy structure formation while dark energy
provides necessary repulsive force for
current accelerated expansion.

Motivated by the desire to explain away the observational fallouts
better and better the form of the equation of state (in short,
EoS) of matter is later generalised in stages first through the
addition of an arbitrary constant with an exponent over the mass
density, generally referred to as generalised Chaplygin gas
(GCG)~\cite{ben}. Barring the serious disqualification
\emph{e.g.}, it violates the time honoured principle of energy
conditions, it is fairly successful to interpret  the
observational results coming out of gravitational lensing or
recent CMBR and SNe data in varied cosmic probes via the fine
tuning  of the value of the newly introduced arbitrary constant.
The form of EoS  is again modified through the addition of an
ordinary mater field, which is termed in the literature as
modified Chaplygin gas (MCG)~\cite{bena,ud,sc2}, claiming an even
better match with
observational results.

 It may be appropriate at this stage to call attention to a recent
 work by Guo and Zhang~\cite{guo1,guo2} where a very generalised
 form of the Chaplygin gas
  relation is invoked, assuming the constant $B$ to depend on the scale
 factor $a$. Taking , \emph{e.g.},$B = B_{0}a^{-n}$, it is shown that for a
  very large value of the scale factor the model interpolates between
 a dust-dominated phase and a quiessence phase (i.e., dark energy with
 a constant equation of state ) ~\cite{dp1,dp} given
 by $ \mathcal{W} = -1 + \frac{n}{6}$.

As mentioned earlier, although different variants of dark energy
models as well as Chaplygin type of gas models are brought in the
cosmological arena as also its associated success to explain the
observations coming out of different cosmic probes it has not
escaped our notice that scant attention has been paid so far  to
address the important issue if all the so called perfect gas models
are atleast thermodynamically stable. Otherwise they would lose
the claim to be treated as a physically realistic system.
Following this there has been of late a resurgence of interests
among workers to address their queries to this aspect of the
problem. Recently Santos et al have studied the thermodynamical
stability in generalised ~\cite{san1} and modified
Chaplygin~\cite{san2} gas model on the basis of standard prescription ~\cite{landau} where both (i)
$\left(\frac{\partial P}{\partial V} \right )_S < 0$,
$\left(\frac{\partial P}{\partial V} \right )_T < 0$  and (ii)
$c_{V} > 0$ are satisfied simultaneously.

 In the present work we attempt to generalise their
results to  Gou-Zhang formalism as also to clarify the issues
concerning thermodynamics of the Chaplygin gas. We also
 investigate the nature of physical parameters such as
 pressure ($P$), effective equation of state ($\mathcal{W}$),
 deceleration parameter ($q$) etc. on the basis of
 thermodynamics.  Later many thermal
quantities are derived as functions of  temperature or volume. In
this case, we also show as consistency check that the third law of
thermodynamics is satisfied by the new form of the Chaplygin gas.
For the generalised Chaplygin gas we expect to have almost similar
behaviour as the Chaplygin gas equations did show. Further, we see
that the Chaplygin gas which shows a unified picture of dark
matter and energy  cools down through the adiabatic expansion of
the universe without any critical point. Returning to the
stability criterion of the Chaplygin gas we find a striking
difference from the analysis of Santos et al. In our case of
variable modified Chaplygin gas (VMCG)~\cite{ud1} we find that the stability
depends critically on the new parameter  $n$, as introduced by Guo
et al. We also notice that stability decreases with the increase
of magnitude of $n$, which apparently disfavors the formalism of
Guo-Zhang~\cite{guo1}. In the section on acoustic velocity we
interestingly note that unlike the previous Chaplygin gas models
where the squared sound velocity is always  positive definite,
the velocity here depends on the value of the newly introduced
parameter $n$. But as noted earlier the stability criteria
dictates that $n$ should be negative which, however, makes the
squared velocity also negative at the late stage of evolution.
This means that the perfect fluid model for Guo's variant of the
chaplygin gas is classically unstable and is in line with the results
obtained earlier by Myung ~\cite{my} while dealing with the
holographic interpretation for Chaplygin gas and tachyon. The
paper ends with a short discussion.

\section*{2. Formalism}
The line element corresponding to spatially flat FRW spacetime is given by

\begin{equation}\label{eq:eq0}
ds^2 = dt^2 - a^2(t) \left(dr^2 + r^2 d \theta^2 + r^2 sin^2 \theta d \phi^2 \right),
\end{equation}
where $a (t)$ is the scale factor.
In this work we consider the following equation of state

\begin{equation}\label{eq:eq1}
P = A \rho - \frac{B}{\rho^{\alpha} }.
\end{equation}
$A$ and $B$ are positive constants. As discussed in the introduction
we have taken $B = B_{0}V^{-\frac{n}{3}}$ where $n$ is an arbitrary constant and $B_{0}$
an absolute constant.
 Here $P$ corresponds to the pressure and $\rho$  the energy density of that fluid such that

\begin{equation}\label{eq:eq2}
\rho = \frac{U}{V},
\end{equation}
where $U$ and $V$ are the internal energy and volume filled by the
fluid respectively.

We try to find out the energy density $U$ and pressure $P$ of Variable
Modified Chaplygin gas as a function of its entropy $S$ and volume
$V$. From general thermodynamics, one has the following
relationship
\begin{equation} \label{eq:eq3}
\left(\frac{\partial U }{\partial V}\right)_{S}= - P.
\end{equation}
With the help of the above equations we get

\begin{equation}\label{eq:eq4}
\left(\frac{\partial U }{\partial V}\right)_{S}=
B_{0}V^{-\frac{n}{3}} \frac{V^{\alpha}}{U^{\alpha}}-A \frac{U}{V},
\end{equation}
which, on integration yields

\begin{equation} \label{eq:eq5}
U = \left[ \frac{3B_{0}(1+\alpha)V^{\frac{3(1+\alpha)-n}{3}}}
{3(A+1)(1+\alpha)-n} + \frac{c}{V^{A(1+\alpha)}}
\right]^{\frac{1}{1+\alpha}}.
\end{equation}
The parameter $c$ is the integration constant, which may be a
universal constant or a function of entropy $S$ only.
 Now we rewrite the above
equation in the following form

\begin{equation}\label{eq:eq6}
U  = \left[\frac{B_{0}(1+\alpha)V^{-\frac{n}{3}}}{N}\right]^{\frac{1}{1+\alpha}}
V \left[1+
\left(\frac{\epsilon}{V}\right)^N\right]^{\frac{1}{1+\alpha}},
\end{equation}
where $N = \frac{3(A+1)(1+\alpha)-n}{3}  > 1$ for
$(A+1)(1+\alpha)> \frac{n}{3}$  for real $U$ and

\begin{equation}\label{eq:eq7}
\epsilon =
\left[\frac{3(A+1)(1+\alpha)-n}{3B_{0}(1+\alpha)}~c\right]^{\frac{1}{N}}
= \left[\frac{Nc}{B_{0}(1+\alpha)}\right]^{\frac{1}{N}},
\end{equation}
which has a dimension of volume. Now the energy density $\rho$ of
the VMCG comes out to be

\begin{subequations}\label{eq:eq8}
\begin{align}
\rho &= \left[\frac{B_{0}(1+\alpha)V^{-\frac{n}{3}}}{N} + V ^{-N - \frac{n}{3}} c\right]^{\frac{1}{1+\alpha}}  \label{eq:eq8a}\\
&= \left[\frac{B_{0}(1+\alpha)V^{-\frac{n}{3}}}{N}\right]^{\frac{1}{1+\alpha}}
 \left[1+
\left(\frac{\epsilon}{V}\right)^N\right]^{\frac{1}{1+\alpha}}  \label{eq:eq8b}.
\end{align}
\end{subequations}
From what has been  discussed above we like to obtain the expression of relevant physical quantities and investigate their behaviour.\\

\textbf{(a) Pressure :}\\

Using equations  \eqref{eq:eq1} and  \eqref{eq:eq8b} the pressure $P$ of the VMCG  may also be determined as a function of
entropy $S$ and volume $V$  in the following form

\begin{eqnarray}\label{eq:eq9}
P =-\left[\frac{B_{0}(1+\alpha)V^{-\frac{n}{3}}}{N}\right]^{\frac{1}{1+\alpha}}
\left(\frac{N}{1+\alpha}\right) \frac{\left[1- \frac{A(1+\alpha)}{N} \left\{1+
\left(\frac{\epsilon}{V}\right)^N \right \}\right]}{\left[1+
\left(\frac{\epsilon}{V}\right)^N\right]^{\frac{\alpha}{1+\alpha}}}.
\end{eqnarray}\\
The equation \eqref{eq:eq9} gives a very general expression of
pressure. Under special conditions

i) For $n=0$ \& $A = 0$, the equation \eqref{eq:eq9}   reduces to
\begin{equation}\label{eq:eq91}
 P = - \frac{(B_{0})^{\frac{1}{1+\alpha}}}{\left \{1+\left(\frac{c}{B_0 V}\right)^{1+\alpha} \right \}^{\frac{\alpha}{1+\alpha} }},
\end{equation}
which is, as expected, similar to generalised Chaplygin gas (GCG)
model ~\cite{san1}.\\

ii) For $n=0$ \& $A \neq 0$, we get the modified Chaplygin gas (MCG) model with pressure given by ~\cite{san2}
\begin{eqnarray}\label{eq:eq92}
P  = - \left(\frac{B_0}{1+A}\right)^{\frac{1}{1+\alpha}} \frac{\left \{ 1+A - A\left[1+ \left\{ \frac{(1+A) c}{B_0V} \right \}^{(1+A)(1+\alpha)}\right] \right \}}{\left[1+\left \{\frac{(1+A) c}{B_0V}  \right \}^{(1+A)(1+\alpha)}\right]^{\frac{\alpha}{1+\alpha}}}.
 \end{eqnarray}\\

iii) For $A = 0$, but $n \neq 0$, the model represents the Variable Chaplygin gas
 (VCGC)~\cite{lu} and equation  \eqref{eq:eq9} becomes

 \begin{equation}\label{eq:eq9a}
 P = - \left(B_{0} V^{-\frac{n}{3}} \right)^{\frac{1}{1+\alpha}}
 \left[\frac{N}{(1+ \alpha) \left \{1 + \left(\frac{\epsilon}{V} \right)^{N}\right\}} \right]^{\frac{\alpha}{1+\alpha}}.
\end{equation}
Again if we put $ \alpha = 1$ in equation \eqref{eq:eq9a}, we get
 $ P = - \left[ \frac{N B_0 V^{-\frac{n}{3}}}{2 \left\{1 + \left(\frac{ \epsilon}{V}
 \right)^N\right\}}\right]^{\frac{1}{2}}$ which is identical with our previous
 work ~\cite{dp2} when we consider the Variable Chaplygin gas (VCG) model.

The pressure may have positive or negative values, depending on the magnitude of both $A$  $\&$
$n$ and also on volume $V$ (fig - 1). For $P=0$, let $V$ is
denoted by $V_{c}$, which is given by

\begin{figure}[ht]
\centering \subfigure[ The graphs clearly show that pressure $P$ can
be both positive and negative depending on the value of $A$.
  For $A = 0$, $P$ is always negative, i.e. Chaplygin type gas with
  $n$ which is also in accord with the equation \eqref{eq:eq1}.]{
\includegraphics[width= 5.8 cm]{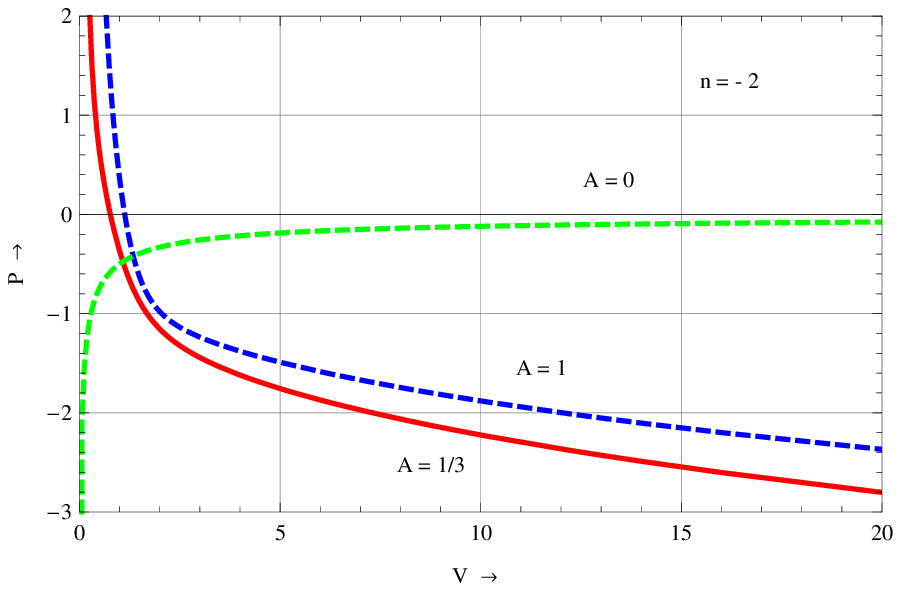}
\label{fig:subfig1} } ~~~~~\subfigure[ It  show that
pressure $P$ is positive  for small value of $V$
  and negative  for large $V$ when $A >0$. As the value of $n$ tends towards
  negative, $P$ becomes  more negative. ]{
\includegraphics[width= 5.8 cm]{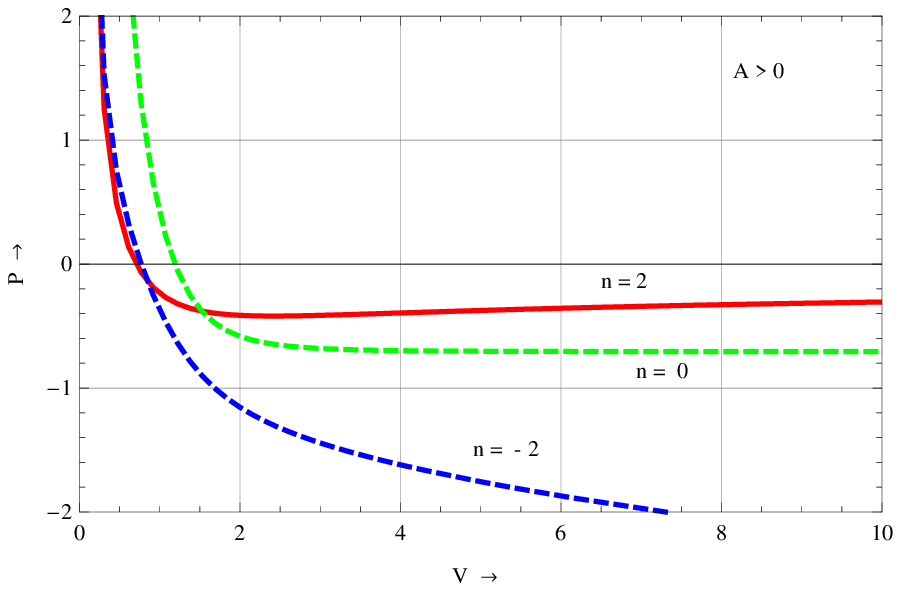}
 \label{fig:subfig2} } \subfigure[For $A = 0$, $n = 0$, $P$ is
  always negative,  chaplygin type case.]{
\includegraphics[width= 6 cm]{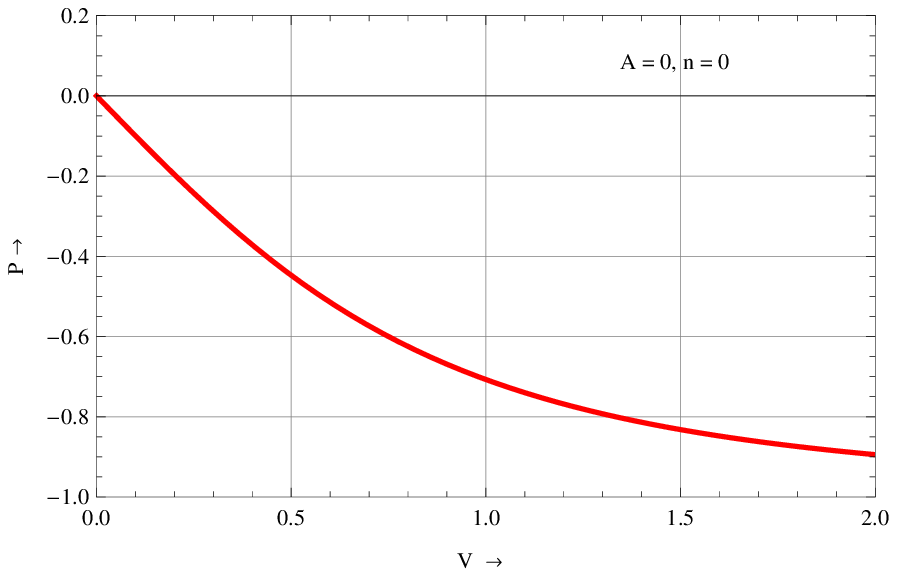}
\label{fig:subfig3} } \label{fig:subfigureExample} \caption[Optional
caption for list of figures]{\emph{The variation of $P$ and $V$  for
different
  values of $A$  and $n$. Here we have taken $B_{0} =1$, $\alpha = 1$ \& $c =1$ for constant $S$.}}
\end{figure}

\begin{equation}\label{eq:eq10}
V_{c} = \epsilon \left[\frac{3A(1+\alpha)}{3(1+\alpha)-n}
\right]^{\frac{1}{N}},
\end{equation}
which  restricts  $n$ as $n < 3(1+
\alpha)$. Initially, \emph{i.e.}, $V < V_{c}$, $P$ is positive,
which indicates a radiation dominated universe. For $V =
V_{c}$, $P = 0$ and $V > V_{c}$, $P$ is negative pointing to a state of  accelerating universe. This is an interesting result showing
 that $V_{c}$ introduces a new scale in the analysis, beyond which a
 dust dominated universe enters the acceleration era.
With expansion an initially decelerating universe tends to reverse
 its motion and prepares to accelerate when its volume crosses a critical value
 designated by $V_{c}$. It is also to be noted that for physically realistic values of constants both $V_{c} $ and $\epsilon$ are of the same order of magnitude i.e. $\epsilon$ also signifies a volume scale beyond which accelerating era commences. So $V \gg \epsilon$ represents a very large volume and $V \ll \epsilon$ the reverse. In what follows we will see that this statement has significant cosmological implications.\\

\textbf{(b)  Caloric EoS:}\\

Now using the expression   \eqref{eq:eq8b} and  \eqref{eq:eq9} we get the caloric
equation of state parameter
\begin{equation}\label{eq:eq11}
\mathcal{W }= \frac{P}{\rho} = A - \left(\frac{N}{1+ \alpha}
\right) \frac{1}{1 + \left(\frac{\epsilon}{V}\right)^N } .
\end{equation}

\begin{figure}[ht]
\begin{center}
   \includegraphics[width=8cm]{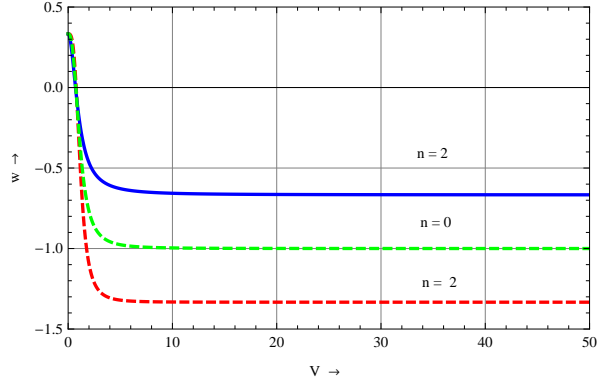}
     \caption{
  \small\emph{
  The variation of $\mathcal{W }$ and $V$  for different
  values of $n$. The values of constants are taken as  $A = \frac{1}{3}, B_{0} =1$, $\alpha = 1$, $c =1$.
     }\label{1}
    }
\end{center}
\end{figure}

As the last EoS is very involved in nature it is very
difficult to extract much physics out of it. So we look forward to
its extremal cases as:

1. For small volume, $V \ll \epsilon$, we get from equation  \eqref{eq:eq11} that

\begin{equation}\label{eq:eq12a}
 P \approx A \rho .
 \end{equation}

  This is a barotropic equation of state. In this case no influence
of $n$ on small volume.

2. For large volume, $V \gg \epsilon$,  the equation \eqref{eq:eq11} reduces to

\begin{equation}\label{eq:eq13}
\mathcal{W }\approx -1 + \frac{n}{3(1+\alpha)}.
\end{equation}
 Equation \eqref{eq:eq10} shows that  $n < 3(1+ \alpha)$. So
  there are three possibilities for  $\mathcal{W }$ depending on the
  signature of $n$ as
  (i) $n > 0$, $ \mathcal{W } > -1$, here the caloric equation of state results in
a quiescence type  and big rip is avoided in this case, (ii) $n = 0$, $\mathcal{W } = -1$, \emph{i.e.},
   we get $\Lambda$CDM.
  (iii) $n <0$, $\mathcal{W } < -1$ represents phantom like universe.
  This is compatible with recent observational results ~\cite{joh}.
  Influence of $n$ is prominent in
 this case.

Initially, \emph{i.e.}, when volume is sufficiently small,
$\mathcal{W }> 0$. As $V$ increases to $V_{c}$, $\mathcal{W }$ tends to
zero. When $V = V_{c} = \epsilon
\left[\frac{3A(1+\alpha)}{3(1+\alpha)-n} \right]^{\frac{1}{N}}$,
$\mathcal{W } = 0$. Again $V$ increases and $\mathcal{W }$
becomes negative. We have seen from fig-2 that $n = 0$,
$\mathcal{W } = -1$, \emph{i.e.}, $\Lambda$CDM model which is
currently fashionable. But for $n > 0$ we get $ 0> \mathcal{W }> -1$ for
large $V$. Similar result was shown  earlier in higher dimensional case
also ~\cite{dp1}. We do not discuss much about $n
> 0$ because later we will show from thermodynamical stability
considerations that the value of $n \leq 0$.\\

 \textbf{(c) Deceleration parameter:}\\

 Now using equation \eqref{eq:eq11} we calculate the deceleration parameter of the VMCG fluid as
\begin{equation}\label{eq:eq14}
q = \frac{1}{2} + \frac{3}{2} \frac{P}{\rho} = \frac{1}{2} +
\frac{3}{2}\left\{A - \left(\frac{N}{1+ \alpha} \right) \frac{1}{1
+ \left(\frac{\epsilon}{V}\right)^N }\right\} ,
\end{equation}

\begin{figure}[ht]
\begin{center}

   \includegraphics[width=8cm]{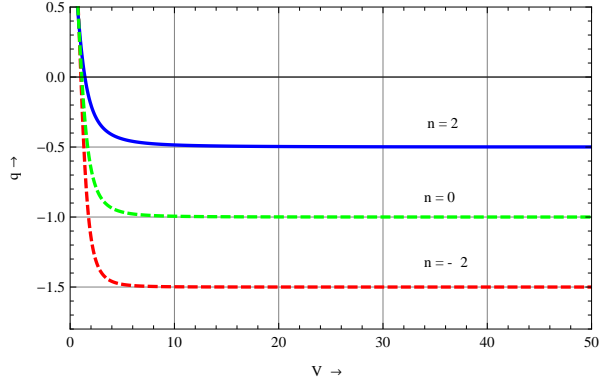}
     \caption{
  \small\emph{
  The variation of $q$ and $V$  for different
  values of $n$. We have considered here $A = \frac{1}{3}, B_{0} =1$, $\alpha = 1$ \& $c =1$.
     }\label{1}
    }
\end{center}
\end{figure}

For mathematical simplicity here also we discuss the extreme
cases.

1. For small volume, $V \ll \epsilon$,
 it gives

\begin{equation}\label{eq:eq15}
q\approx  \frac{1}{2} + \frac{3}{2}A ,
\end{equation}
\emph{i.e.}, $q$ is positive, universe decelerates for small
$V$.\\
2.  For large volume, $V \gg \epsilon$, the equation \eqref{eq:eq14} reduces to

\begin{equation}\label{eq:eq16}
q\approx -1 + \frac{n}{2(1+\alpha)} .
\end{equation}
Initially, \emph{i.e.}, when volume is very small there is no
effect of $n$  on $q$, here $q$  is positive, universe decelerates. But
for large volume,  $q$ is negative and  this  depends on the value of
$n$ also. For \emph{flip} in velocity to occur the flip volume
($V_{f}$) becomes

\begin{equation}\label{eq:eq17}
V_{f} = \epsilon \left[\frac{(1+3A)(1+
\alpha)}{2(1+\alpha)-n}\right]^{\frac{1}{N}}.
\end{equation}
Therefore $n < 2(1+ \alpha)$, which interestingly does not violate
our previous restriction on $n$. A little analysis of the above
equation shows that for $V_{f}$ to have real value $n < 2(1+
\alpha)$, otherwise there will be no \emph{flip}. This also
follows from the fig - 3 where only $n <4$ allows \emph{flip} (for $\alpha =1)$.
Alternatively the inequality $n < 2(1+ \alpha)$ may be treated as
our acceleration condition. Thus $V < V_{f}$, we get deceleration; and $V >
V_{f}$,  acceleration occur.

It has not also escaped our notice that we are here getting two
scales ($V_c$ and $V_f$) for volume where apparently acceleration
flip occurs. Again we know that in a FRW
cosmology for flip to occur pressure should not only be negative
but its magnitude should be less than $\frac{\rho}{3}$
(\emph{i.e.}, $\rho + 3 P <0 $ ) as well. So in our case one
should get $V_{c} < V_{f}$, which also
follows from their expressions \eqref{eq:eq10} and \eqref{eq:eq17}.
 In this connection it may also be mentioned that with $V = V_{f}$, $\rho + 3P = 0$ as expected.\\

\textbf{(d) Velocity of Sound:}\\

Let us consider $v_s$ be the velocity of sound,  then using equation
 \eqref{eq:eq9} we can write

 \begin{equation}\label{eq:eq34}
v_{s}^2= \left(\frac {\partial P}{ \partial \rho} \right)_S =  A +
\frac{N \alpha}{ (1 + \alpha) \left \{1 + \left(\frac{\epsilon}{V}
\right)^N\right \} }  - \frac{n N}{n + (3N + n) \left(\frac{\epsilon}{V}\right)^N}.
 \end{equation}

 Since sound  speed should be   $0< \left(\frac {\partial P}{ \partial \rho}
  \right)_S < 1$, now our analysis shows ,

1. For small volume, \emph{i.e.}, at early universe, the equation \eqref{eq:eq34}
 leads to $ 0 < A < 1$, and this limit includes $A = \frac{1}{3}$,
 the radiation dominated universe.

2. For large volume, the equation \eqref{eq:eq34} reduces to

\begin{equation}\label{eq:eq35}
  v_{s}^2 =  - 1 + \frac{n}{3(1+ \alpha)}.
 \end{equation}

The equation \eqref{eq:eq35} does not depend on $A$, it depends on $n$ and
 $\alpha$ only. In what follows we shall see that from the thermodynsmical stability condition   the value of
 $n < 0$,  leading to a phantom universe ~\cite{joh}. Moreover the equation  \eqref{eq:eq35}
  gives an imaginary speed of sound for $\alpha > 0$, leading to a perturbative cosmology. One need not be too sceptic about it because  it favours structure formation~ \cite{fab}.

  It may not be out of place to draw some correspondence to a recent work
  by Y.S. Myung ~\cite{my} where a comparison is made
  between holographic dark energy, Chaplygin gas, and tachyon model with constant potential.
We  know that their squared speeds are crucially important to
determine the stability of perturbations. They found that the
squared speed for holographic dark energy is always negative when
imposing the future event horizon as the IR cutoff, while those
for original Chaplygin gas and tachyon are always non-negative.
This is in sharp contrast to our variant (VMCG) of the Chaplygin
gas model where we observe that depending on the signature of $n$
the squared velocity may be both positive or negative. However as
discussed earlier a non negative value of $n$ is clearly ruled out
from thermodynamic stability considerations. This points to the
fact that the perfect fluid model for VMCG  dark energy model is
classically unstable like the holographic interpretation for
Chaplygin gas and tachyon and hence problematic in the long run.\\

\textbf{(e) Thermodynamical Stability:}\\

 To verify the thermodynamic stability conditions of a
fluid along its evolution, it is necessary \emph{(a)} to determine if the
pressure reduces both for an adiabatic  and isothermal expansion~\cite{landau} $\left(
\frac{\partial P}{\partial V}\right)_{S} < 0$  $\&$ $\left(
\frac{\partial P}{\partial V}\right)_{T} < 0$
 and \emph{(b)} also to examine
if the thermal capacity at constant volume, $c_{V} >0$.

 Using equations  \eqref{eq:eq1} and  \eqref{eq:eq9} we get

\begin{eqnarray}\label{eq:eq18}
\left( \frac{\partial P}{\partial V}\right)_{S}
=\frac{P}{3V(1 + \alpha)}\frac{ A(1+\alpha)  \left\{n \left[1 +
\left(\frac{\epsilon}{V} \right)^N \right]
 + 3N\left(\frac{\epsilon}{V} \right)^N \right \}
  + N \left\{\frac{3N \alpha \left(\frac{\epsilon}{V} \right)^N}{\left[
1+ \left(\frac{\epsilon}{V} \right)^N \right]} - n  \right\}  }{N- A
(1+ \alpha)\left[1 +
\left(\frac{\epsilon}{V} \right)^N \right]}.~
\end{eqnarray}

\begin{figure}[ht]
\begin{center}
   \includegraphics[width=8cm]{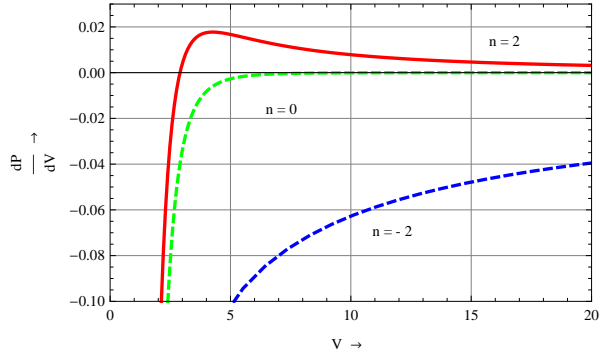}
     \caption{
  \small\emph{
  The variation of $\left(\frac{\partial P}{\partial V}\right)_{S}$ and
  $V$  for different
  values of $n$. The nature of graphs shows that for  $n \leq 0$,
   $\left(\frac{\partial P}{\partial V}\right)_{S} < 0$ throughout the
    evolution unlike $n>0$ where $\left(\frac{\partial P}{\partial V}\right)_{S}
     <0$ only at the early stage. We have taken  $A = 1, B_{0} =1$, $\alpha = 1$, $c =1$.
     }\label{1}
    }
\end{center}
\end{figure}

Now we have to examine the negativity of $\left(\frac{\partial P}{\partial V}
\right)_{S}$. As this expression  \eqref{eq:eq18} is so involved   we can
not make any conclusion considering this equation as a whole. Our analyses naturally
split into two parts :
\begin{enumerate}
  \item Firstly, we have considered small volume, $V  \ll \epsilon$,   where  equation \eqref{eq:eq18}  gives $ \left(\frac{\partial P}
{\partial V}\right)_{S}\approx - (1+A)\frac{P}{V}$. We get from previous
analysis  that at the early stage of
evolution $P \approx A \rho$, therefore, in this case $\left(\frac
{\partial P}{\partial V}\right)_{S} \approx -A(1+A) \frac{ \rho}{V}$
 which is independent of $n$ but very much depends on $A$ and at this stage of evolution
 $\left( \frac{\partial P}{\partial V} \right)_S < 0$ .

  \item For large volume, $V \gg \epsilon$, the equation \eqref{eq:eq18}
    reduces to $\left( \frac{\partial P}{\partial V} \right)_S \approx -
     \frac{n P}{3 V (1 + \alpha)}$. Since $P$ is negative at the late
     stage of evolution, so $n$ must be negative to make $\left
     ( \frac{\partial P}{\partial V} \right)_S < 0$. We have seen that the
     dependence of $n$ is prominent at the later case. From fig - 4 we get the similar type of conclusion.
\end{enumerate}

Now we discuss some special cases to constrain  the parameters used here.

(i) The simultaneous conditions $ n =0$, $\alpha = 0$ and
 $A = 0$ must be discarded because it will place a severe
 restriction on the stability of this fluid, in such a case $\left(\frac
 {\partial P}{\partial V} \right)_S =0$
  and the pressure will remain the same through any adiabatic change
   of volume. However $B_{0}$ here behaves like a Cosmological Constant.
    We get the de-Sitter type of metric.

  (ii) Again for $\alpha = 0$, $A =0$ and $n \neq 0$,
    $ \left( \frac{\partial P}{\partial V} \right)_S = \frac{n}{3}
     B_0 V^{-(1+\frac{n}{3})} $, \emph{i.e.},
    for $ n < 0$, $ \left(\frac{\partial P}{\partial V} \right)_S < 0$ which
    is in disagreement with the previous work of
    Santos \emph{et al} ~\cite{san1} where simultaneously  $\alpha = 0$, $A =0$  is not possible.
     Relevant to mention that,  $n \leq 0$ implies that the pressure goes more and more
      negative with volume. This agrees with the observational results~\cite{guo1}.

(iii) When $A = 0$, $n = 0$ and $ \alpha \neq 0$, the  equation \eqref{eq:eq18}
 reduces to Santos's work
~\cite{san1}. In this case   $ \left( \frac{\partial P}
{\partial V} \right)_S$ will be

\begin{eqnarray}\label{eq:eq18a}
 \left( \frac{\partial P}{\partial
V}\right)_{S}  =  \alpha \frac{ P}{V}
\left(\frac{\epsilon}{V} \right)^{1+\alpha} \left \{ 1+ \left(\frac{\epsilon}{V}
 \right )^{1 + \alpha} \right\}^{ - 1},
 \end{eqnarray}
 which gives  $ \left( \frac{\partial P}{\partial
V}\right)_{S}   < 0 $ for $\alpha >0$. It also agrees with the work of Santos
\emph{et al}~\cite{san1} in this field.

 (iv)  when  $\alpha = 0$, $n = 0$ and $ A \neq 0$,  $\left( \frac{\partial P}{\partial
V}\right)_{S} = - \frac{ A B_0}{V} \left(\frac{\epsilon}{V} \right)$ , \emph{i.e.},
$\left( \frac{\partial P}{\partial
V}\right)_{S}  < 0 $ which implies simultaneously $A > 0$ and $ B_0 >0$.

 (v) For $A \neq 0$ but $\alpha \neq 0$ \& $n = 0$ the equation  \eqref{eq:eq18}
   reduces to

\begin{eqnarray}\label{eq:eq18b}
\left( \frac{\partial P}{\partial V}\right)_{S} =  \frac{P}{V} \frac{\left
( \frac{\epsilon }{V}\right)^N}{1 - \frac{A}{1+A} \left \{ 1 +  \left
(\frac{\epsilon }{V}\right)^N \right \}} \left[A + \frac{ \alpha
(1+\alpha)}{1 +  \left(\frac{\epsilon }{V}\right)^N} \right].
\end{eqnarray}
This equation \eqref{eq:eq18b} is identical with the another work of Santos
\emph{ et al} ~ \cite{san2}.

(vi) Again when  $A =0$ but $\alpha \neq 0$ \& $n \neq 0$ the equation
\eqref{eq:eq18} gives

\begin{eqnarray}\label{eq:eq19}
\left( \frac{\partial P}{\partial V}\right)_{S} =  \frac{P}{3V (1+\alpha)}
\left[ \frac{3N \alpha\left( \frac{\epsilon }{V}\right)^N}{  1 +  \left(\frac
{\epsilon }{V}\right)^N } - n \right].
\end{eqnarray}
This is the case of Variable Generalised Chaplygin gas (VGCG) model~\cite{lu}.
\begin{figure}[ht]
\begin{center}
   \includegraphics[width=8cm]{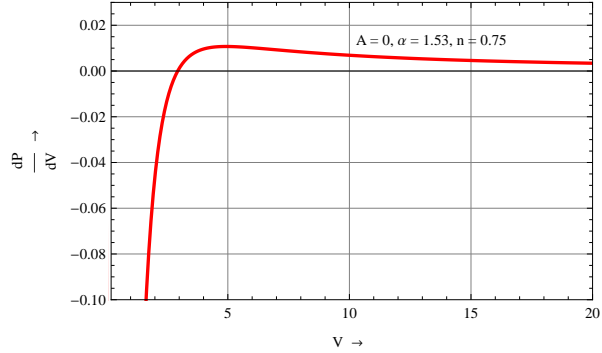}
     \caption{
  \small\emph{
  The variation of $\left( \frac{\partial P}{\partial V}\right)_{S}$ and
   $V$   is shown for $A = 0$, $ \alpha = 1.53$, $ n = 0.75$ and also $ B_{0} =1$, $c =1$.
     }\label{1}
    }
\end{center}
\end{figure}

It is clear from equation  \eqref{eq:eq19}  that for $\alpha > 0$, $n$
should be negative for $\left( \frac{\partial P}{\partial V}\right)_{S} < 0$
throughout the evolution.  It may not be out of place to call attention
 to an earlier work of Lu~\cite{lu} where he studied the Union SNe Ia data
  and Sloan Digital Sky Survey(SDSS) baryon acoustic peak to  constrain
  the Variable Generalised Chaplygin Gas (VGCG) model where he obtained the
   best fit values of $n = 0.75$ and $\alpha = 1.53$. Our analysis shows
   that Lu's conclusion is untenable if consideration of thermodynamical
    stability is taken into account. Similar conclusion may be drawn from
     the analysis of our fig-5. In this connection it may be pointed out that
      we have studied ~\cite{dp2} for the case of $A =0$ but $\alpha = 1$
       \& $n \neq 0$, where we have seen that $\left( \frac{\partial P}
       {\partial V}\right)_{S}<0$  for $n < 0$ throughout the evolution.

But our analysis is based on  all these constants:
 $A$, $B_0$, $\alpha$ and $n$. As it is not possible to constrain  all the parameters simultaneously from equation \eqref{eq:eq18}, we have tried it step by step. Now we want to concentrate on  the signature of $n$ rather  than the other parameters because we are dealing with VMCG model where $n$ has a special role throughout the evolution, particularly  at late time.  Now $n =0$  was  discussed earlier by several authors~\cite{san2}. It is seen from the equation  \eqref{eq:eq18} that for positive values of  $A$, $B_0$ and $\alpha$, it will be  $\left( \frac{\partial P}{\partial
V}\right)_{S} <0 $ when $n < 0$ throughout the evolution. Fig-4 gives similar type of conclusion. Again it is seen from the fig-4 that for the positive value of $n$ at small volume $\left( \frac{\partial P}{\partial
V}\right)_{S} <0$ but as volume increases, \emph{i.e.}, at the late universe where the influence of $n$ is significant,  for  $n > 0$, $\left( \frac{\partial P}{\partial V}\right)_{S} > 0$; so the stability  is questionable  throughout the evolution for $n >0$.  Therefore, in the context of thermodynamical stability, we have to conclude that  the value of $n$ should be negative for the VMCG model.

Now we have to examine if $\left( \frac{\partial P}{\partial V}\right)_{T} \leq 0 $  as well. We will show in the next section that for $n < 0$ this condition may also be satisfied.

 One should also verify the positivity of thermal capacity
at constant volume $c_{v}$  where
   $c_{v}=T \left(\frac{\partial S }{\partial T} \right)_{V} =
    \left(\frac{\partial U }{\partial T} \right)_{V}=V \left(\frac{\partial \rho }{\partial T} \right)_{V}$. Now we determine the temperature $T$ of the Variable modified Chaplygin gas as a
    function of it volume $V$ and its entropy $S$. The temperature
    $T$ of this fluid is determined from the relation  $T = \left(\frac{\partial U }{\partial S}
    \right)_{V}$. Using the above relation of the
    temperature and with the help of equation  \eqref{eq:eq5} we get the
    expression of $T$ as

\begin{subequations}\label{eq:eq20}
\begin{align}
T &=
\frac{V^{1-N-\frac{n}{3(1+\alpha)}}}{1+\alpha}\left[\frac{B_{0}(1+\alpha)}{N}
+ V^{-N}c \right]^{-\frac{\alpha}{1+\alpha}}\left( \frac{\partial c}{\partial S} \right)_{V}   \label{eq:eq20a} \\
&= \frac{V^{1-N-\frac{n}{3}}}{1+\alpha}\rho^{- \alpha} \left( \frac{\partial c}{\partial S} \right)_{V}  \label{eq:eq20b}.
\end{align}
\end{subequations}

If $c$ is also assumed to be a universal constant, then
$\frac{dc}{dS}=0$ and the fluid, in such a condition, remains at
zero temperature for any value of its volume and pressure.
Therefore, to discuss extensively the thermodynamic stability of
the variable modified Chaplygin gas whose temperature varies
during its expansion, it is necessary to assume that the
derivative of equation  \eqref{eq:eq20} is not zero implying  $ \left(\frac{\partial c }{\partial S}
\right) \neq 0$.  We have no \emph{a priori} knowledge of the functional
dependence of $c$. From physical considerations, however,
 we know that this function must be such as to give
positive temperature and cooling along an adiabatic expansion and
so we choose that $\left(\frac{\partial c }{\partial S}
\right)>0$.

Now from dimensional analysis, we observe from equation \eqref{eq:eq5} that

 \begin{equation}\label{eq:eq21}
 \left[U \right] = \left\{\frac{\left[c \right]}{\left[V \right
 ]^{A(1+\alpha)}} \right\}^{\frac{1}{1+ \alpha}}.
\end{equation}
Since $\left[U \right] = \left[T\right]\left[S \right]$,  we can write

 \begin{equation}\label{eq:eq22}
 \left[c \right] = \left[T \right]^{1+ \alpha}\left[S \right]^{1+ \alpha}\left[V \right]^{A(1+ \alpha)}.
\end{equation}
It is difficult to get an analytic solution of $c$ from equation \eqref{eq:eq22}, so as a trial case, we take an empirical expression of $c$ and then to check  if the resulting expressions satisfy standard relations of thermodynamics.
But as $c$ is a function of entropy only,  the expression of $c$ will be

\begin{equation}\label{eq:eq23}
 c = \left( \tau v^A \right)^{1+ \alpha} S^{1+ \alpha},
\end{equation}
where $\tau$ and $v$ are constants having the dimensions of time and volume respectively.\\
Now

\begin{equation}\label{eq:eq24}
\frac{dc}{dS} = (1+ \alpha) \left( \tau v^A \right)^{1+ \alpha} S^{\alpha}.
\end{equation}
Using equation \eqref{eq:eq20} and \eqref{eq:eq24}, we get the expression of temperature

\begin{subequations}\label{eq:eq25}
\begin{align}
T &=
V^{1-N-\frac{n}{3(1+\alpha)}}\left[\frac{B_{0}(1+\alpha)}{N}
+ V^{-N}c \right]^{-\frac{\alpha}{1+\alpha}}\left( \tau v^A \right)^{1+\alpha}S^{\alpha}   \label{eq:eq25a} \\
&= V^{1-N-\frac{n}{3}} \left( \tau v^A \right)^{1+ \alpha} \rho^{- \alpha} S^{\alpha} \label{eq:eq25b}\\
&=\frac{\tau v^A}{V^A} \left \{ 1- \frac{1}{1+ \left(\frac{\epsilon}{V} \right)^N} \right \}^{\frac{\alpha}{1+ \alpha}},   \label{eq:eq25c}
\end{align}
\end{subequations}
and from equation \eqref{eq:eq25a},  the entropy is

\begin{equation}\label{eq:eq26}
S = \frac{\left[\frac{B_{0}(1+\alpha)}{N} V^{N} \right]^{\frac{1}{1+\alpha}} \left(\frac{T}{\tau^{1+\alpha}} \right)^{\frac{1}{\alpha}} \left(\frac{V}{v^{1+\alpha)}} \right)^{\frac{A}{\alpha}}}{\left \{1-\left(\frac{T}{\tau} \right)^{\frac{1+\alpha}{\alpha}}\left(\frac{V}{v} \right)^{\frac{A(1+\alpha)}{\alpha}}\right \}^{\frac{1}{1+\alpha}}}~,
\end{equation}

It follows from  equation \eqref{eq:eq26} that for positive and finite entropy one should have $0 < TV^A < \tau v^A$, but individually $0 < T< \tau $ and $v < V < \infty$, \emph{i.e.}, $\tau$ represents the maximum temperature whereas $v$ represents the minimum volume. It is shown from the equation \eqref{eq:eq25c} that as volume of the VMCG increases, temperature decreases. It is also proved from the equation \eqref{eq:eq25c} that when $T \rightarrow 0$, $V \rightarrow \infty$ and when   $T \rightarrow \tau$, $V \rightarrow v$. Thus we can apparently avoid the initial singularity.

Evidently  at $T = 0$, $S = 0$ which implies that the third law of thermodynamics is satisfied in this case.

Now using equation  \eqref{eq:eq26}
 we get the expression of thermal heat capacity as

\begin{equation}\label{eq:eq27}
c_{V}  = T\left(\frac{\partial S}{\partial T} \right)_{V} =\frac{\left[\frac{B_{0}(1+\alpha)}{N} V^{N} \right]^{\frac{1}{1+\alpha}} \left(\frac{T}{\tau^{1+\alpha}} \right)^{\frac{1}{\alpha}} \left(\frac{V}{v^{1+\alpha}} \right)^{\frac{A}{\alpha}}}{ \alpha \left \{1-\left(\frac{T}{\tau} \right)^{\frac{1+\alpha}{\alpha}}\left(\frac{V}{v} \right)^{\frac{A(1+\alpha)}{\alpha}}\right \}^{\frac{2+ \alpha}{1+\alpha}}}~.
\end{equation}

Since $0 < TV^A < \tau v^A$ and $\alpha > 0$,  $c_V >0$ is always satisfied irrespective of the value of $n$. This ensures the positivity of $\alpha$.
It is interesting to note that when the temperature goes to zero $c_V$ goes
 to zero as expected from the third law of thermodynamics.

If we put $A =0$ and $\alpha = 1$, \emph{i.e.} Variable  Chaplygin gas model, we get the identical expression of $c_{V}$ of  our previous work ~\cite{dp2}. Again for $A =0$ and $n = 0$, the equation \eqref{eq:eq27} reduces to the work of Santos et al~\cite{san1}.

    To end the section a final remark may be in order. While
\emph{positivity} of specific heat is strongly desirable
\emph{vis-a-vis} when dealing with special relativity, in a recent
communication Luongo \emph{et al} ~\cite{luo} argued that in a FRW type
of model like the one we are discussing  a negative specific
heat at constant volume and a vanishingly small  specific heat at
constant pressure $(c_P)$ are compatible with observational data. In fact
they have derived the most general cosmological model which is
agreeable  with the $c_{V}< 0$ and $c_{P}\sim 0$ values obtained for
the specific heats of the universe and showed, in addition, that  it
also overcomes the fine-tuning and
the coincidence problems of the $\Lambda$CDM model.\\

\textbf{(f) Thermal EoS:}\\

Since $P = P(V,T)$, using \eqref{eq:eq5}, \eqref{eq:eq23} and \eqref{eq:eq26} we get the internal energy as a function of both $V$ and $T$ as follows:

\begin{equation}\label{eq:eq28}
  U= V \left \{ \frac{\frac{B_{0}(1+\alpha)}{N} V^{-\frac{n}{3}}}{ {  1-\left(\frac{T}{\tau} \right)^{\frac{1+\alpha}{\alpha}}\left(\frac{V}{v} \right)^{\frac{A(1+\alpha)}{\alpha}}}} \right\}^{\frac{1}{1+ \alpha}}.
\end{equation}
Now using \eqref{eq:eq1}, \eqref{eq:eq2} and \eqref{eq:eq28}   the Pressure is

\begin{equation}\label{eq:eq29}
P  = \rho \left[ A - \frac{N}{1+ \alpha}  \left \{{  1-\left(\frac{T}{\tau} \right)^{\frac{1+\alpha}{\alpha}}\left(\frac{V}{v} \right)^{\frac{A(1+\alpha)}{\alpha}}} \right \} \right ].
\end{equation}

We have seen from equation \eqref{eq:eq29} that for $A=0$ and $\alpha = 1$, the solution reduces to our previous work~\cite{dp2}. It is to be seen that for $A =0$ and $n = 0$ the above solution goes to an earlier work of Santos et al \cite{san1}.

Now the thermal EoS parameter is given by

\begin{equation}\label{eq:eq30}
 \omega = A - \frac{N}{1+ \alpha}  \left \{{  1-\left(\frac{T}{\tau} \right)^{\frac{1+\alpha}{\alpha}}\left(\frac{V}{v} \right)^{\frac{A(1+\alpha)}{\alpha}}} \right \}.
\end{equation}

Thus thermal EoS parameter is, in general,  a function of both temperature and volume. If we consider  early stage of the universe when at very high temperature and small volume, \emph{i.e.}, at $T \rightarrow \tau$ and also   $V \rightarrow v$,  we get from equation \eqref{eq:eq30} that $\omega \approx A$, \emph{i.e.}, $P \approx A \rho$. This is same as  equation \eqref{eq:eq12a}.

\begin{figure}[ht]
\begin{center}
   \includegraphics[width=8cm]{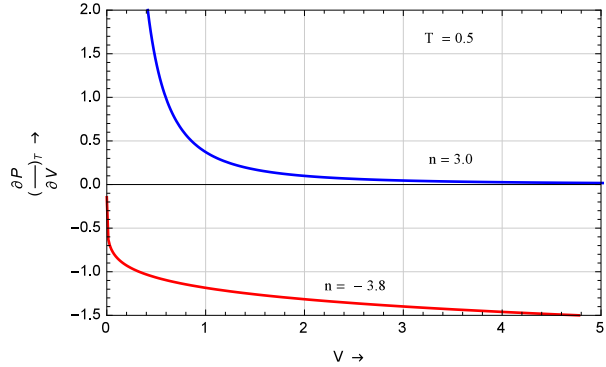}
     \caption{
  \small\emph{ The variations of $\left(\frac{\partial P}{\partial V}\right)_{T}$ vs
   $V$   are shown. The graphs clearly show that $\left(\frac{\partial P}{\partial V}\right)_{T} <0$ throughout the evolution only for negative values of $n$.
   Here we have taken  $A = 0.1$, $ \alpha = 0.1$, $B_{0} = 1$,  $\tau = 1$ and $v = 1$.
       }\label{1}
    }
\end{center}
\end{figure}

Secondly we consider for large volume,\emph{ i.e.}, very low temperature , \emph{i.e.}, $T \rightarrow 0$, we get from equation \eqref{eq:eq30} that $\omega \approx -1 + \frac{n}{3(1+ \alpha)}$, which is identical with equation \eqref{eq:eq17} as is customary with the existing literature in this field ~\cite{san1,dp2}. Thus thermodynamical state represented by equations \eqref{eq:eq12a} and \eqref{eq:eq13} are essentially same at both early and the late stage of the universe.

Now from equation \eqref{eq:eq29} we have to examine if $\left(\frac {\partial P}{\partial V} \right)_{T} \leq 0 $, and a lengthy but straight forward calculation shows  that  only  for negative value of $n$, this condition satisfies. It is very difficult to find out $\left(\frac{ \partial P}{ \partial V} \right)_{T} $ from the equation \eqref{eq:eq29} in a compact form and to get requisite inferences from it. So we have taken recourse to graphical method instead. From fig - 6, we clearly infer that $\left(\frac{ \partial P}{ \partial V} \right)_{T} < 0$ for $n <0$  throughout the evolution depending on the values of constants.

Now to discuss the thermodynamic stability of the VMCG one should have $\left(\frac{ \partial P}{ \partial V} \right)_{T} < 0$ and $\left(\frac{ \partial P}{ \partial V} \right)_{S} < 0$ \emph{i.e.}, both isothermal and adiabatic situations need to be addressed. In the process we have found that $\left(\frac{ \partial P}{ \partial V} \right)_{S} $ and  $\left(\frac{ \partial P}{ \partial V} \right)_{T} $  are negative for  $n < 0$. Relevant to mention that in a previous  work by Santos et al ~\cite{san2} it was assumed $\left(\frac{ \partial P}{ \partial V} \right)_{T} = 0$ for simplicity, but in our case we have not made such a simplistic approach. In fact we have used a different type of approach to find out the expression of $c = c(S)$.  As we have not  made this assumption our analysis is more general in nature and our formulations would not reduce to there case when $n = 0$.

As pointed out in the previous section we here take up a standard relation of thermodynamics  (e.g. the first internal energy equation)~\cite{zym} as a trial case to see if the equation \eqref{eq:eq23} is satisfied.

\begin{equation}\label{eq:eq30a}
 \left(\frac{\partial U}{\partial V} \right)_{T} = T  \left(\frac{\partial P}{\partial T} \right)_{V} -P.
\end{equation}

Using equations \eqref{eq:eq28} and \eqref{eq:eq29}  we verify the  relation \eqref{eq:eq30a} which proves the correctness of our approach.\\

\textbf{(f) Pressure-Volume relation:}\\

It is very difficult to arrive at a $(P\sim V)$ relation in a general way. So one has to take recourse to extremal conditions.

 1. For   small volume,  $V  \ll \epsilon$,  which gives $P \sim A \rho$. In this case energy density  and also pressure are very high. Using  \eqref{eq:eq25c} the temperature will be

 \begin{equation}\label{eq:eq31}
T \approx \frac{ \tau v^A}{V^A}.
 \end{equation}

At the early stage of the universe $V \rightarrow v$ (minimum volume) implies that $T \rightarrow \tau$ (maximum temperature)
So the temperature is high enough at this stage. Using equations  \eqref{eq:eq25c}  and  \eqref{eq:eq31} we get

 \begin{equation}\label{eq:eq32}
\rho \approx S  \frac{ \tau v^A}{V^{1+A}},
 \end{equation}
 therefore,
   \begin{equation}\label{eq:eq33}
 U V^A = \rho V^{A+1}= S \tau v^A.
  \end{equation}
        We know at this stage that $P \sim A \rho = A \frac{U}{V}$, \emph{i.e.}, $PV \sim AU $,  which gives using equation \eqref{eq:eq33} $P V ^{1+A} = S \tau v^A$. Since the entropy remains constant in an adiabatic process, this relation leads to $P V ^{1+A} = $ Constant. So it is observed that for small volume, \emph{i.e.}, at high temperature  the VMCG behaves as a fluid of  $\gamma (=\frac{c_P}{c_V}) = 1+A$. We can also rewrite the EoS  as $ P = \left( \gamma -1 \right) \rho$. Since early universe is radiation dominated \emph{i.e.}, $A = \frac{1}{3}$, the value of $\gamma = 1+A = \frac{4}{3}$, the pressure is related to the volume as $PV^{\frac{4}{3}}= $ constant. Thus the VMCG behaves like a photon gas. The equation of adiabatic photon gas coincides with extreme relativistic electron gas.

 2. For large volume, $V >> \epsilon$,

    Due to low density at this stage,  entropy density is sufficiently small at the low temperature.
    We know from equation \eqref{eq:eq13} that
    \begin{equation}\label{eq:eq32b}
    P \approx \left\{ -1 + \frac{n}{3(1+ \alpha)} \right\} \rho.
     \end{equation}
    Again, from equation \eqref{eq:eq8b}, in this case we get

    \begin{equation}\label{eq:eq32b1}
    \rho \approx \left\{ \frac  {B_{0} (1+\alpha)V^{-\frac{n}{3}}}{N} \right\}^{\frac{1}{1+\alpha}}.
    \end{equation}
    Now using equations \eqref{eq:eq32b} and \eqref{eq:eq32b1}, we get

    \begin{equation}\label{eq:eq32c}
    PV^{\frac{n}{3(1+\alpha)}} =  \left \{ -1 + \frac{n}{3(1+\alpha)} \right \}\left\{ \frac  {B_{0} (1+\alpha)}{N} \right\}^{\frac{1}{1+\alpha}}.
    \end{equation}
    Equations  \eqref{eq:eq32b},  \eqref{eq:eq32b1} and  \eqref{eq:eq32c} can be obtained from the equations  \eqref{eq:eq30},  \eqref{eq:eq28} and  \eqref{eq:eq29} respectively at $T \rightarrow 0$. For an adiabatic system, the above equation looks like $PV^{\gamma} =  k$, where $k$ is a constant. Since we know that VMCG is thermodynamically stable for  $n < 0$ which leads to $k < 0$. In this case, $P$ is also negative. At the late stage of evolution, \emph{i.e.}, at low temperature, the VMCG behaves like a fluid of $\gamma  = \frac{n}{3(1+\alpha)}$. Interestingly, it is to be noted that the value of this  $\gamma$  depends on  both $n$ and $\alpha$ at this stage of evolution.  It further follows from equation  \eqref{eq:eq25c} that $T \rightarrow 0$  for $\frac{\epsilon}{V} << 1$. Since any system near $T = 0$ is in states very close to its ground state, quantum mechanics is essential to the understanding of its properties. Indeed, the degree of randomness at these low temperatures is so small that discrete quantum effects can be observed on a macroscopic scale.

\section*{3. Discussion}

 Following the discovery of late acceleration of the
universe there is a proliferation of varied dark energy models as
its possible rescuers. While many of them significantly explain
the observational findings coming out of different cosmic probes
serious considerations have not been directed so far to the query
whether the models are thermodynamically viable, for example if
they obey the time honoured stability criteria. We have here
considered a very general type of exotic fluid, termed `Variable
Modified Chaplygin gas'  and studied its cosmological implications, mainly its thermodynamical stability.
 Regarding the cosmological dynamics we have come across two
 characteristic volumes of the fluid  - $V_c$ and $V_f$
 representing critical volume and flip volume respectively.
 The former refers   to the case where pressure changes its sign while
 the latter gives the volume when the acceleration flip occurs. From
 physical consideration one should get $V_c < V_f$ -  which also
 matches with our analysis. As discussed at the end of section  $2 (c)$, for
 flip to occur in FRW cosmology only a negative pressure is not the
 necessary and sufficient condition. The magnitude of pressure
 should be also less than $\frac{\rho}{3}$, which also follows from our analysis.\\

Although the exhaustive analysis of the latest cosmological
observations provides a definite clue of the existence of dark
energy in the universe but it is difficult to distinguish between
the merits of various forms of dark energy at present. For the stability criteria
we have followed the standard prescription :
 $\left(\frac{ \partial P}{ \partial V} \right)_{S} < 0$,
 $\left(\frac{ \partial P}{ \partial V} \right)_{T} < 0$ and
  $c_{V} > 0 $. This, however, dictates that the new parameter, $n$
  introduced VMCG should be negative definite.  This contrasts sharply
  with an earlier contention of Lu's~\cite{lu}
  where to explain the observational results they
have to choose $n$ as  positive definite. This is pathological
because it makes the system thermodynamically unstable.

Again this model shows that at early stage, the EoS becomes
$P = A \rho$, where $ 0 \leq A \leq 1$. But at late stage,
it reduces to equation \eqref{eq:eq32b}.
From the thermodynamical stability conditions, we find that   $ n < 0$,
 which favours a  phantom like evolution and big rip is thus unavoidable. So
far as phantom model is concerned, it is found to be compatible
with SNe Ia observations and CMB anisotropy measurements ~\cite{joh}.
 The most important
 conclusion coming out of our analysis is that this model covers
  both big bang and big rip in the whole evolution process.

  It is to be noted that at $T = 0$ the entropy of VMCG
  vanishes as in conformity  with the third law of thermodynamics.
  We have  studied both the thermal and the caloric EoS which shows
   that both $0 < T < \tau$ and $v < V < \infty $ where  $\tau$ is a maximum
    temperature and $v$ is a minimum volume attainable. Here $\tau$ and $v$
     are canonical in the sense that as $T \rightarrow \tau$, $V \rightarrow v$
      at the early stage, \emph{i.e.}, $\tau $ represents the maximum temperature
      that our VMCG model can sustain for the small finite volume, $v$.

      We have also discussed  $(P \sim V)$ relation and at early stage ($A = \frac{1}{3}$)
  it is shown that for an isentropic system VMCG behaves
  like a photon gas, as it was at the time of radiation dominated era.
  It is also shown that for large volume,\emph{ i.e.}, at low temperature, entropy
   is sufficiently low which agrees well with the currently available
   low energy density of the universe.

   Finally it may not be out of place to point out that as the field equations are very
   involved in nature we have to adopt an \emph{ansatz} to determine the function of integration $c = c(S)$.
   However to justify it we have checked an important relation  \eqref{eq:eq30a} (first internal energy equation)  and have found it to be correct.
   So our \emph{ansatz} is essentially viable.  Moreover, this model is very  general
   in the sense that many of earlier works in this field may be obtained
   as a special case.

   To end a final remark may be in order. We have here concentrated on the
   cosmological  and thermodynamical behaviour of the VMCG model
    mainly on a theoretical premise. However, as a future exercise one should try to constrain the value of the parameters associated with both thermal and caloric EoS in the light of observational values.

 \vspace{0.5cm}

\textbf{Acknowledgment : } One of us(SC) acknowledges the
financial support of UGC, New Delhi for a MRP award and
acknowledges CERN for a short visit. DP acknowledges the financial support of
UGC, ERO for a MRP (No- F-PSW- 165/13-14) and also acknowledges IRC,NBU for short visit. The authors wish to thank the anonymous referee for valuable comments and suggestions.

\end{document}